\documentclass[aps,prl,twocolumn,superscriptaddress,showpacs]{revtex4}
\usepackage{amsfonts}
\usepackage{amssymb}
\usepackage{amsmath}
\usepackage{graphicx}
\usepackage{dcolumn}
\usepackage{bm}
\usepackage{flafter}

\begin{document}

\title{Breakdown of the interlayer coherence in twisted bilayer graphene}

\author{Youngwook Kim}
\affiliation{Department of Physics, Pohang University of Science and Technology, Pohang 790-784,  Korea}
\author{Hoyeol Yun}
\affiliation{Division of Quantum Phases and Devices, School of Physics, Konkuk University, Seoul, 143-701, Korea}
\author{Seung-Geol Nam}
\affiliation{Department of Physics, Pohang University of Science and Technology, Pohang 790-784,  Korea}
\author{Minhyeok Son}
\affiliation{Department of Chemistry and Division of Advanced Materials Science, Pohang University of Science and Technology, Pohang 790-784, Korea}
\author{Dong Su Lee}
\affiliation{Institute of Advanced Composite Materials, Korea Institute of Science and Technology, Wanju-gun 565-902, Korea}
\author{Dong Chul Kim}
\affiliation{Department of Electronics and Telecommunications, Norwegian University of Science and Technology, NO-7491 Trondheim, Norway}
\author{S. Seo}
\affiliation{Department of Physics, Sejong University, Seoul, 143-747, Korea}
\author{Hee Cheul Choi}
\affiliation{Department of Chemistry and Division of Advanced Materials Science, Pohang University of Science and Technology, Pohang 790-784, Korea}
\author{Hu-Jong Lee}
\affiliation{Department of Physics, Pohang University of Science and Technology, Pohang 790-784,  Korea}
\author{Sang Wook Lee}
\email{leesw@konkuk.ac.kr}
\affiliation{Division of Quantum Phases and Devices, School of Physics, Konkuk University, Seoul, 143-701, Korea}
\author{Jun Sung Kim}
\email{js.kim@postech.ac.kr}
\affiliation{Department of Physics, Pohang University of Science and Technology, Pohang 790-784,  Korea}

\date{\today}

\begin{abstract}
Coherent motion of the electrons in the Bloch states is one of the fundamental concepts of the charge conduction in solid state physics. In layered materials, however, such a condition often breaks down for the interlayer conduction, when the interlayer coupling is significantly reduced by e.g. large interlayer separation. We report that complete suppression of coherent conduction is realized even in an atomic length scale of layer separation in twisted bilayer graphene. The interlayer resistivity of twisted bilayer graphene is much higher than the c-axis resistivity of Bernal-stacked graphite, and exhibits strong dependence on temperature as well as on external electric fields. These results suggest that the graphene layers are significantly decoupled by rotation and incoherent conduction is a main transport channel between the layers of twisted bilayer graphene.

\end{abstract}


\maketitle

In many layered systems, the interlayer coupling is one of the key parameters for altering their electronic properties~\cite{inter:cuprates,inter:organic,inter:chalcogenide,caxispress:uher:review,fieldcaxisgraphite:kopelevich:review,superlattice:osada:inter}. When a thick insulating block is inserted between the metallic layers, the interlayer coupling can be significantly reduced, leading to breakdown of the interlayer coherence, as nicely demonstrated in two dimensional electron gas (2DEG) in semiconductor superlattice\cite{superlattice:osada:inter}. Such an "interlayer version" of the Mott-Ioffe-Regel limit is realized when the layer separation exceeds the mean free path across the layers, which is evidenced by qualitatively different temperature dependences of the intralayer (metallic) and the interlayer (semiconducting) resistivities. The intriguing interlayer conduction has been observed in various systems including high-$T_c$ cuprates~\cite{inter:cuprates}, organic crystals~\cite{inter:organic}, dichalcogenides~\cite{inter:chalcogenide}, graphite~\cite{caxispress:uher:review,fieldcaxisgraphite:kopelevich:review} and semiconductor superlattices~\cite{superlattice:osada:inter}, but its underlying mechanism related to the interlayer incoherence has been under debate in last decades~\cite{inter:maslov}.

Graphene, an ideal 2DEG system, also exhibits rich electronic properties depending on how it is stacked on top of another graphene layer~\cite{reviewpaper:geim:review,Moireband:macdonald:review}. While bilayer graphene in Bernal stacking has massive charge carriers with a zero band gap, twisted bilayer graphene with a random orientation of the layers has a massless electronic dispersion similar to that of monolayer graphene~\cite{twistelectronic:castroneto:review}. Twisted bilayer graphene is of particular interest because several intriguing properties such as renormalization of Fermi velocity~\cite{twistelectronic:castroneto:review,localize:trambly:review}, van Hove singularities\cite{vhs:luican:review}, and electronic localization~\cite{localize:trambly:review,twistquantumhall:DSlee:paper} were recently discovered. Experimental studies including angle-resolved photoemission spectroscopy~\cite{arpes:de heer:review}, scanning tunneling spectroscopy~\cite{vhs:luican:review}, Raman spectroscopy~\cite{reductionfremiv:Ni:review,raman:zettle:paper}, and in-plane transport~\cite{decouplesdoapl:schimidt:review,twistmit:pablo:paper}, suggest that the layers are decoupled in twisted bilayer graphene, and the misoriented layers are often considered as being electrically isolated. However, it is still not clear what sense the layers are decoupled on an atomic length scale of the layer separation, and how strong the interlayer coupling is in twisted bilayer graphene~\cite{Moireband:macdonald:review,transporttwist:macdonald:review}.

In this Letter, we present experimental evidence for complete suppression of the interlayer coherence between in twisted bilayer graphene. We found that the interlayer resistivity of twisted bilayer graphene is much higher than that of Bernal-stacked graphite by at least 4 orders of magnitude. In contrast to the almost temperature-independent intralayer resistivity, the interlayer resistivity exhibits strong negative temperature dependence. These results demonstrate that $incoherent$ tunneling is a main conduction channel between the layers, due to the momentum-mismatch of the 2 dimensional Fermi surfaces from each layer. We thus suggest that the twisted bilayer graphene provides a rare example of the layered systems showing the interlayer incoherence even with a small layer separation of $\sim$ 4 ${\rm \AA}$.

\begin{figure}
\begin{center}
\leavevmode
\includegraphics[bb=56 400 340 686, width=8.5cm]{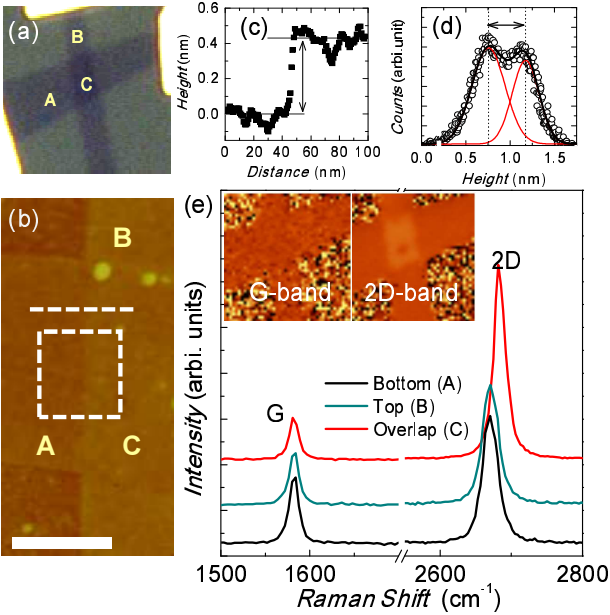}
\caption{\label{fig1} (color online) (a) Optical image of a cross junction of two monolayer graphene sheets (A and B) forming twisted bilayer graphene in the overlapped region (C). (b) AFM image with a scale bar of 500 nm. (c) The height profile along the line indicated in (b). (d) The height histogram of the boxed region in (b). The data is fitted by the sum (black) of two Gaussian functions (red), yielding $d$ $\approx$ 0.4 nm as indicated by the arrow. (h) Raman spectra taken from the monolayer (A and B) and the twisted bilayer (C) regions. The inset shows the Raman map of the relative frequency shift of the G band and the 2D band in the twisted bilayer region with respect to those of the monolayer region.}
\end{center}
\end{figure}

In order to fabricate the graphene cross junction we employed a mechanical transfer process~\cite{supp}. Twisted bilayer graphene is formed in the overlapped region of two monolayer graphene layers, as shown in Fig. 1(a). During the process, the surfaces of both the lower and the upper graphene layers are kept free of contact to chemical solvents until they overlap with each other. This is crucial for ensuring tight contact between the layers without any chemical adsorbates in-between.

Before discussing the interlayer transport properties of twisted bilayer graphene, we confirm that in the overlapped region, two graphene layers are tightly stacked. Figure 1 shows the optical and the atomic force microscopy (AFM) images of the twisted bilayer graphene sample that is mainly discussed in this manuscript. The separation between the layers is $d$ $\approx$ 0.4~nm, determined by the line-cut height profile~[Fig. 1(c)] and the height histogram over the step from the bottom to the top layers~[Fig.~1(d)]. This is very close to the layer spacing of $d$ $\sim$~0.34~nm in Bernal-stacked graphite.

Twisted stacking of the graphene layers in the overlapped region is also demonstrated by Raman spectroscopy. According to recent studies~\cite{reductionfremiv:Ni:review,raman:zettle:paper}, the double
resonance peak~(2D) of twisted bilayer graphene resembles a single Lorentzian peak rather than
multiple peaks as found in Bernal-stacked bilayer graphene. Furthermore, the Fermi velocity
of twisted bilayer graphene is known to be reduced as compared to the monolayer inducing a blue-shift
of the 2D peak.
Both features are found in our sample when we compare the spectra of the overlapped region~(C) with
those of the bottom~(A) and the top~(B) layers. The Raman 2D peak at C is blue-shifted (2682~cm$^{-1}$) as compared to the peaks at A and B (2672~cm$^{-1}$). By contrast, the Raman G peaks appear at the same position of $\sim$ 1582~cm$^{-1}$ for the twisted bilayer and the monolayer regions~[Fig.~1(e)]. These observations confirm that the graphene layers are indeed coupled at high energy levels in the overlapped region. According to Ref.~[\onlinecite{raman:zettle:paper}], a blue-shift of the 2D peak by $\sim$~10~cm$^{-1}$ with an enhancement of its intensity corresponds to the twist angle of 15$^{\rm o}$-30$^{\rm o}$.

\begin{figure}
\begin{center}
\leavevmode
\includegraphics[bb=40 330 455 770, width=8.5 cm]{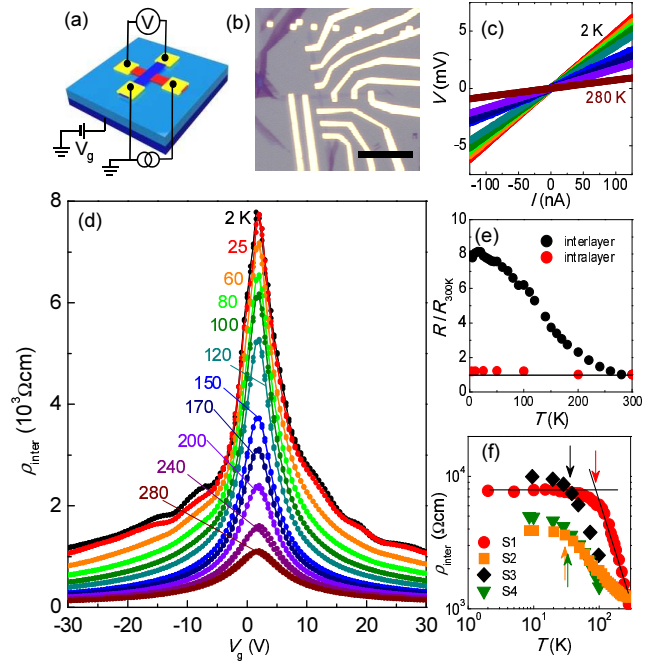}
\caption{\label{fig2} (color online) (a) Schematic diagram of a twisted bilayer graphene device. (b) Optical image of a twisted bilayer graphene device (S1) with a scale bar of 10 $\mu$m. (c) Current-voltage characteristics of the interlayer conduction at various temperatures from 2 K to 280 K (using the same color code in (d)) (d) The interlayer resistivity of twisted bilayer graphene (S1) with variation of the gate voltages ($V_g$) at different temperatures. (e) The temperature dependence of the normalized interlayer and the intralayer resistivity at $V_g$ = $V_N$. (f) Temperature dependence of the $\rho_{inter}(T)$ for several twisted bilayer graphene devices (S1-S4) at $V_g$ = $V_N$. The crossover temperature $T^*$ from the $T$-independent to the strong $T$-dependent regimes is indicated by the arrows.}
\end{center}
\end{figure}

The interlayer transport measurements in the twisted layers were done using four branches of monolayers as electrical leads. For both bottom and top monolayers, the intralayer resistivity as a function of back-gate voltage~($V_g$) shows a narrow peak at $V_g$ = 2 V without significant temperature dependence~\cite{intra}. For the interlayer transport, we apply the current between each branch of the bottom and the top layers and measure the interlayer voltage drop using the other two branches as illustrated in Fig.~2(a). In total, four devices of the twisted bilayer graphene were studied (Table 1), and all devices show qualitatively similar behaviors of the interlayer conduction.  We present the results obtained from S1~[Fig.~2(b)] unless otherwise noted. Figure~2(c) displays the current-voltage ($I$-$V$) characteristics through the layers at the charge neutral point with a back-gate voltage of $V_g$=$V_N$ in the temperature range from 2~K to 280~K. The $I$-$V$ characteristics exhibit the ohmic behaviors, and it is sustained down to 2~K.

Despite the linear $I$-$V$ characteristics the interlayer resistivity (${\rho}_{inter}$) is very large. Figure~2(d) shows the
interlayer resistivity $\rho_{inter}$ as a function of $V_g$ at various temperatures. Here we define
${\rho}_{inter} \equiv R_{inter}{\cdot}A/d$, where $A$ is the area of the overlapped region, $d$ is the interlayer distance, and $R_{\mathrm{inter}}$ is the interlayer resistance\cite{supp}.
At 280 K, ${\rho}_{\mathrm{inter}} {\sim}$ 2000 ${\Omega}$cm, which is at least four orders of magnitude higher than
the reported values of the $c$-axis resistivity ${\rho}_c$ for graphite. Typically ${\rho}_{c}$ ranges from $\sim 10^{-4}$ ${\Omega}$cm for single crystalline graphite~\cite{caxispress:uher:review} to ${\sim} 10^{-1}$ ${\Omega}$cm for highly oriented pyrolytic graphite (HOPG)~\cite{caxispress:uher:review,fieldcaxisgraphite:kopelevich:review}.

\begin{table}
  \caption{The characteristics of twisted bilayer graphene samples. The interlayer resistivity ($\rho_{inter}$), the interlayer equilibration rate ($\tau_{RC}^{-1}$) at 2 K are listed together with the crossover temperature ($T^*$) for $V_g$ = $V_N$. The relative shift of 2D Raman band ($\Delta_{2D}$) of the twist bilayer graphene with respect to that of monolayer graphene is also listed.}
  \label{tbl:samples}
  \begin{tabular}{ccccc}
    \hline
     Sample & $\rho_{inter}(2 K)$ ($k\Omega$cm) & $\tau_{RC}^{-1}$ (s$^{-1}$) &  $T^*$ (K) & $\Delta_{2D}$ (cm$^{-1})$  \\
    \hline
     S1 & 7.80 & 5.54$\times$10$^8$ & $\sim$ 90 & $\approx$ 8 \\
     S2 & 3.98 & 1.11$\times$10$^{9}$ & $\sim$ 30 & $\approx$ 10 \\
     S3 & 10.00 & 4.33$\times$10$^{8}$ & $\sim$ 42 & \\
     S4 & 5.24 & 8.25$\times$10$^{8}$ & $\sim$ 40 & \\
    \hline
  \end{tabular}
\end{table}

The interlayer resistivity shows a strong temperature dependence as shown in Fig. 2(e). $\rho_{inter}(T)$ at charge neutral point~($V_g$~=~$V_N$) increases with lowering temperature, and  it saturates at $\sim$ 7800 ${\Omega}$cm below the crossover temperature $T^*$ $\sim$ 90 K. The similar temperature dependence was observed in other samples~[Fig. 2(f)]. The negative temperature dependence ($d{\rho}_{inter}/dT$~$<$ 0) is distinct from the intralayer resistance measured for the bottom and the top monolayers, which is almost temperature-independent~[Fig.~2(e)]. More importantly, this behavior contrasts to the metallic behavior, $d{\rho}_c/dT$~$>$~0 found in single crystalline Bernal-stacked graphites~\cite{caxispress:uher:review}. These results provide experimental evidence for the \emph{incoherent} interlayer conduction between the layers~\cite{note_HOPG}.
\begin{figure}
\begin{center}
\leavevmode
\includegraphics[bb=38 275 500 775, width=8.5cm]{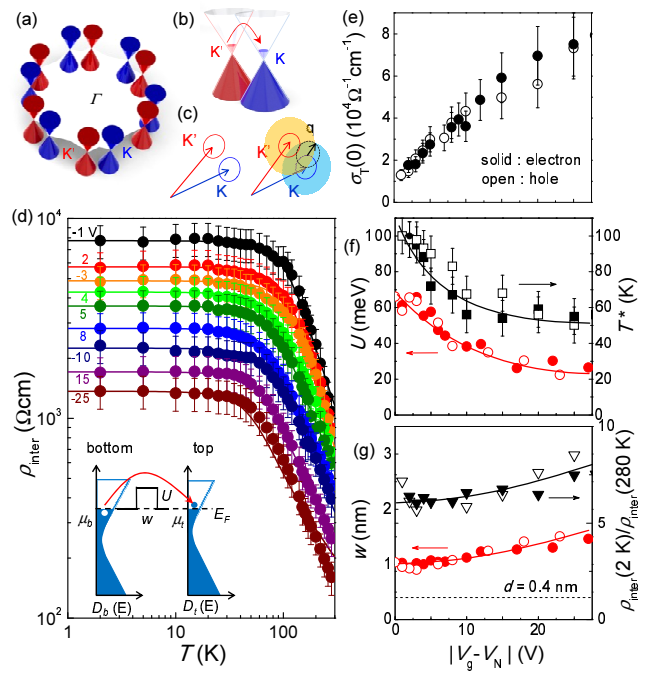}
\caption{\label{fig4} (color online) (a) The momentum mismatch of two sets of Dirac cones (red and blue) from different layers in twisted bilayer graphene. (b) The neighboring Dirac cones overlaps at high energies. Thermally-excited tunneling between two FS's of each layer is indicated by the arrow. (c) Phonon-assisted tunneling with a momentum exchange of $q$ enhances the overlap between the FS's. (d) The interlayer resistivity ($\rho_{inter}$) of twisted bilayer graphene (S1) at different $V_g$'s as a function of temperature. The solid lines are the fit of the metal-insulator-metal junction model whose schematic illustrations are shown in the inset~(see the text). (e) The $V_g$ dependence of the conductivity $\sigma(0)$ at zero temperature limit. The open~(solid) symbols represent the case of hole~(electron) carriers. The height $U$ (f) and the width $w$ (g) of an effective barrier estimated from the fitting are shown as a function of $V_g$. The solid line is the guide-to-eyes. The crossover temperature $T^*$ (f) and the resistivity ratio $\rho_{inter}$(2 K)/$\rho_{inter}$(280 K) (g) are also plotted. The interlayer separation $d$ = 0.4 nm is indicated by the dashed line in (g).}
\end{center}
\end{figure}

Recent theoretical studies on twisted bilayer graphene suggested that Dirac cones from the two monolayers are misoriented as shown in Fig. 3(a)~\cite{transporttwist:macdonald:review}. When the twist angle is away from the commensurate angle, the overlap of the FS's from each layer does not occur, and the energy difference between states with the same extended momentum is typically much larger than the Fermi energy. In this case, the interlayer tunneling of charge carriers with conservation of their in-plane momentum is significantly suppressed. The interlayer equilibriation time $e.g.$ the RC time $\tau_{RC}$ for the interlayer conduction is expected to be 10$^{-6}$ - 10$^{-12}$ sec for $n$ = 5 $\times$ 10$^{12}$ cm$^{-2}$ at zero temperature limit. Using the measured $\sigma(2~{\rm K})$ at $V_g$ = 30 V and the interlayer dielectric constant $\epsilon_{GG}$ $\sim$ 2.6$\epsilon_{0}$~\cite{twistmit:pablo:paper}, we estimate $\tau_{RC}$ $\sim$ 2 $\times$ 10$^{-10}$ sec, that is well within the range of theoretical estimates~\cite{transporttwist:macdonald:review}.

For quantitative analysis, we employ a model of metal-insulator-metal junction~\cite{tunneling:text}. The current through a potential barrier is given by
\begin{equation}
I \sim \int {T(E)D_b(E)D_t(E+eV)[f_b(E)-f_t(E+eV)]dE},
\end{equation}
where $D_{b,t}(E)$ and $f_{b,t}(E)$ are the density of states (DOS), and the Fermi-Dirac distribution function for the bottom and the top monolayers, respectively.
In the model, the conductivity saturates at a zero temperature limit ${\sigma}(0) \sim D_b(E_F)D_t(E_F)T(0)$. Assuming no significant $V_g$-dependence of $T(0)$, ${\sigma}(0)$ $\sim$ $|V_g-V_N|$, which roughly agrees with our experiments [Fig. 3 (e)]. The finite and relatively large ${\sigma}(0)$ near the charge neutral point is due to the electron-hole puddles~\cite{puddle}, where the charge density inhomogeneity of $n_0 \sim 3 \times 10^{11}$~cm$^{-2}$ are expected~\cite{note_DOS}.

We now discuss the temperature dependence of the interlayer resistivity, which is determined by tunneling probability between the two monolayers. Since the interlayer tunneling occurs when the Fermi circles intersect in the extended Brillouin zone\cite{transporttwist:macdonald:review}, it is allowed for excited carriers at higher energies where the two neighboring Dirac cones overlap each other as shown in Fig.~3(b). Thus we can consider an effective barrier with its height $U$ corresponding to the energies required for interlayer tunneling between two misoriented FS's~\cite{note_U}. The temperature dependence of $\rho_{inter}(T)$ is well reproduced by the fit using Eq. (1) as shown in Fig.~3(e). Here we assumed a square barrier with a height of $U$ and a width of $w$~[Fig.~3(e)] using the Wentzel-Kramers-Brillouin approximation. For the DOS of monolayer graphene, the phenomenological DOS is used with a broadening factor $\Gamma$ = 60 meV~\cite{supp,phenoDOS,note_DOS}.

The resulting barrier height $U$ is reduced at high $|V_g-V_N|$'s showing electron-hole symmetry, which resembles the behaviors of $T^*$ [Fig.~3(f)]. This is indeed what is expected since thermal energy needed for the interlayer tunneling decreases when the Fermi circles become larger with increase of carrier density. The barrier width $w$ is found to be 1.0-1.3 nm with a moderate $V_g$ dependence similar to that of $\rho_{inter}(2~{\rm K})/\rho_{inter}(280~{\rm K})$~[Fig.~3(g)]. Note that the estimated $w$ is about 3 times larger than the interlayer separation $d$ $\approx$ 0.4 nm measured by the AFM~[Fig. 1]. We tested various cases with different magnitude and shape of $U$, but  $w$ $>$ $d$ $\approx$ 0.4 nm is needed in order to explain the observed strong temperature dependence, $i.e.$ a large value for the ratio $\rho_{inter}(2~{\rm K})/\rho_{inter}(280~{\rm K})$.

We attribute this discrepancy to the interlayer tunneling assisted by phonon at high temperatures~\cite{inter:maslov}. In this case, the carriers are scattered by acoustic phonon modes within the phonon sphere of diameter $q$ $\approx$ $k_BT/\hbar v_s$ where $v_s$ is sound velocity. In particular, in graphene $q$ can be much larger than the size of Fermi circle at moderate temperatures~\cite{phonon_graphene}, and thus phonon-assisted tunneling significantly enhances the overlap between the misoriented FS's as illustrated in Fig.~3(c).
With lowering temperature, therefore, not only the thermally-excited tunneling but also the phonon-assisted tunneling are suppressed, and the effect of the momentum mismatch of the FS's become pronounced. This results in the stronger temperature dependence of $\rho_{inter}(T)$, leading to more significant suppression of interlayer tunneling than expected in the metal-insulator-metal junction model. Further studies are highly desirable for understanding the incoherent interlayer tunneling process assisted by phonon and other bosons, as discussed recently~\cite{inter:maslov}. Our findings clearly manifest that the momentum-mismatch of the FS's due to misorientation is essential for breakdown of the interlayer coherence in twisted bilayer graphene.

In conclusion, using graphene cross junctions we investigate the interlayer transport properties of twisted bilayer graphene. The large magnitude of the resistivity and its strong dependence on temperature and also on external electric fields reveal that the main interlayer conduction mechanism is incoherent tunneling due to the momentum-mismatch of FS's. Thus the twisted bilayer graphene is a rare example of the layered systems where the interlayer incoherence is induced by misorientation of the layers even with an atomic scale of layer separation. We envisage that the interlayer conduction can be modified by the insertion of other nanosheets using multiple stacking~\cite{vertical:geim:paper} or chemical intercalation~\cite{fecl:philipkim:reveiw}. Therefore, graphene cross junctions with further functionalization can provide a new platform incorporating vertical transport of graphene layers for possible electronic applications.

\begin{acknowledgments}
The authors thank G. T. Kim, M. Jonson, and E. E. B. Campbell for fruitful discussion. This work was supported through BSR (2009-0076700, 2011-0021207      ), EPB Center (2010-0001779), SRC (2011-0030046), and WCU (R31-2008-000-10057-0), CRC (2010K000981) programs by the National Research Foundation of Korea (NRF) funded by the Ministry of Education, Science and Technology.

\end{acknowledgments}


\begin{thebibliography}{00}
\bibitem{intra} The mobilities of electron (hole) carriers are as high as 13400 (11100) cm$^2$/Vs and 8960 (8820) cm$^2$/Vs for the top and bottom layers, respectively.
\bibitem{note_HOPG} In HOPG, similar semiconducting temperature dependence of ${\rho}_{\mathrm{c}}$ has been observed at high temperatures, but it is followed by the metallic behavior at low temperatures~\cite{fieldcaxisgraphite:kopelevich:review,caxispress:uher:review}. This can be due to the heterogeneous conduction of Bernal-stacked domains with stacking faults induced by the in-plane mosaicity. The twisted bilayer graphene, therefore, is well in the incoherent regime, which is an extreme opposite to Bernal-stacked graphite via HOPG.
\bibitem{note_DOS} $n_0$ is estimated from the full-width-half-maximum of the $R_{intra}$($V_g$) curves for the monolayer layer. The corresponding DOS broadening factor near the charge neutral point is $\sim$ 60 meV in the phenomenological model for the DOS of monolayer graphene.
\bibitem{note_U} Since the momemtum-mismatch between the pair of neighboring Dirac cones are different, $U$ should be considered as an effective value averaged over the extended Brillouin zone.
\end{thebibliography}
\end{document}